**Title:** Autonomous Reporting of 'Normal' Chest X-rays by Artificial Intelligence in the United Kingdom; Can We Take the Human Out of the Loop?

**Type of Manuscript:** Opinion

**Authors**

**Dr Katrina Nash*,** MBChB (Hons), BMedSci (Hons)
OxCAIR, Oxford University Hospitals, Oxford, United Kingdom
Address: OxCAIR, Kadoorie Centre, Oxford University Hospitals, Headley Way, OX3 9DU
0000-0002-5204-9688
katrina.nash@ouh.nhs.uk

**Dr James Vaz*,** BMBS, PGDip, PGCert
OxCAIR, Oxford University Hospitals, Oxford, United Kingdom
Address: OxCAIR, Kadoorie Centre, Oxford University Hospitals, Headley Way, OX3 9DU
0000-0002-0513-7220
james.vaz@ouh.nhs.uk

**Dr Ahmed Maiter ,** MB BChir, MA, FRCR
Department of Radiology, Sheffield Teaching Hospitals FT Trust, Sheffield, UK
NIHR Sheffield Biomedical Research Centre, Sheffield, UK
School of Medicine & Population Health, University of Sheffield, Sheffield UK
Address: Radiology department, Sheffield Teaching Hospitals NHS FT Trust, Northern General Hospital, Herries Rd, S5 7AU
0000-0002-4999-2608
ahmed.maiter@nhs.net

**Dr Christopher S Johns,** MBBS, FRCR, PhD
Radiology department, Sheffield Teaching Hospitals NHS FT Trust, Sheffield
Address: Radiology department, Sheffield Teaching Hospitals NHS FT Trust, Northern General Hospital, Herries Rd, S5 7AU
0000-0003-3724-0430
Christopher.johns@nhs.net

**Dr Nicholas Woznita,** MBE, PhD
Imaging Department, UCLH, UK
Address: Imaging Department UCLH 235 Euston Road London, School of Allied & Public Health Professions Canterbury Christ Church University

0000-0001-9598-189X
nicholas.woznitza@nhs.net

**Dr Aditya U Kale**, MBChB
College of Medicine and Health, University of Birmingham, UK
National Institute for Health and Care Research Birmingham Biomedical Research Centre, University of Birmingham, Birmingham, UK
University Hospitals Birmingham NHS Foundation Trust, Birmingham, UK
Warwick Medical School, University of Warwick, Coventry, UK
University Hospitals Coventry and Warwickshire, Coventry, UK
Address: College of Medicine and Health, University of Birmingham, Edgbaston, Birmingham B15 2TT, UK
0000-0002-2186-1446
 a.kale@bham.ac.uk

**Dr Abdala T Espinosa Morgado**, MD, MSc
Oxford Clinical Artificial Intelligence Research (OxCAIR), Oxford University Hospitals, Oxford, UK
Address: OxCAIR, Kadoorie Centre, Oxford University Hospitals, Headley Way, OX3 9DU
0000-0003-0967-3554
abdala.espinosa@ouh.nhs.uk

**Dr Rhidian Bramley**, MBChB MRCP FRCR
The Christie NHS Foundation Trust, Manchester M20 4BX
Clinical Lead for Diagnostics, Digital and Innovation, Greater Manchester Cancer
Address: Address: The Christie NHS Foundation Trust, Manchester M20 4BX
0009-0005-0293-053X
rhidian.bramley@nhs.net

**Dr Mark Hall,** MBChB, FRCR, DipMedEd
Queen Elizabeth University Hospital, Govan, Glasgow
Address: Queen Elizabeth University Hospital, Govan, Glasgow, G51 4TF
0000-0002-6310-0872
mark.hall2@ggc.scot.nhs.uk

**Professor David Lowe**, MBChB MSc MD
Digital Health Validation Lab, University of Glasgow, Glasgow, UK

Emergency Department, Queen Elizabeth University of Hosptial, Glasgow, UK
Address: Queen Elizabeth University Hospital, Govan, Glasgow, G51 4TF
0000-0003-4866-2049
David.lowe@glasgow.ac.uk

**Professor Alex Novak,** MSc, BSc (Hons), MBChB
OxCAIR, Oxford University Hospitals, Oxford, UK
Emergency Medicine Research Oxford (EMROx), Oxford University Hospitals NHS Foundation Trust, Oxford, UK
Address: OxCAIR, Kadoorie Centre, Oxford University Hospitals, Headley Way, OX3 9DU
0000-0002-5880-8235
alex.novak@ouh.nhs.uk

**Dr Sarim Ather,** BSc (Hons), MBChB (Hons), PhD, FRCR
OxCAIR, Oxford University Hospitals, Oxford, UK
Address: OxCAIR, Kadoorie Centre, Oxford University Hospitals, Headley Way, OX3 9DU
0000-0001-9614-5033
sarim.ather@ouhs.nhs.uk

*Joint first author*

**Conflicts of Interest**

SA is a director and shareholder at RAIQC Ltd. AN has previously done consulting work for GE Healthcare Ltd. NW has previously done consulting work for InHealth, Apollo Radiology International, and SMR Healthtech, and received travel grant from Qure.ai 2023. AM has received reimbursements from Bayer plc as part of the British Institute of Radiology AI Fellowship. AM is a member of the Clinical Radiology AI Faculty for the Royal College of Radiologists. DL has received grant funding from qure.ai and Annalise.ai. There are no other conflicts of interest to declare.

**Funding**

NW has received funding from SBRI Healthcare, HEE/NHSE London.

AM is funded by the NIHR Sheffield Biomedical Research Centre (BRC). The views expressed are those of the author(s) and not necessarily those of the NHS, the NIHR or the Department of Health and Social Care (DHSC).

**Abstract**

Chest X-rays (CXRs) are the most commonly performed imaging investigation. In the UK, many centres experience reporting delays due to radiologist workforce shortages. Artificial intelligence (AI) tools capable of distinguishing "normal" from "abnormal" CXRs have emerged as a potential solution. If "normal" CXRs could be safely identified and reported without human input, a substantial portion of radiology workload could be reduced.

This article examines the feasibility and implications of autonomous AI reporting of "normal" CXRs. Key issues include defining "normal," ensuring generalisability across populations, and managing the sensitivity-specificity trade-off. It also addresses legal and regulatory challenges, such as compliance with IR(ME)R and GDPR, and the lack accountability frameworks for errors. Further considerations include the impact on radiologists practice, the need for robust post-market surveillance, and incorporation of patient perspectives. While the benefits are clear, adoption must be cautious, with strong governance, legal clarity, and rigorous clinical validation to ensure safe and sustainable use.

# Autonomous Reporting of 'Normal' Chest X-rays by Artificial Intelligence in the United Kingdom; Can We Take the Human Out of the Loop?

## Abstract

Chest X-rays (CXRs) are the most commonly performed imaging investigation. In the UK, many centres experience reporting delays due to radiologist workforce shortages. Artificial intelligence (AI) tools capable of distinguishing "normal" from "abnormal" CXRs have emerged as a potential solution. If "normal" CXRs could be safely identified and reported without human input, a substantial portion of radiology workload could be reduced.

This article examines the feasibility and implications of autonomous AI reporting of "normal" CXRs. Key issues include defining "normal," ensuring generalisability across populations, and managing the sensitivity-specificity trade-off. It also addresses legal and regulatory challenges, such as compliance with IR(ME)R and GDPR, and the lack accountability frameworks for errors. Further considerations include the impact on radiologists practice, the need for robust post-market surveillance, and incorporation of patient perspectives. While the benefits are clear, adoption must be cautious, with strong governance, legal clarity, and rigorous clinical validation to ensure safe and sustainable use.

**Why autonomous reporting?**

Chest X-rays (CXRs) account for around 40% of all diagnostic imaging.[1] Despite their clinical importance, timely reporting of CXRs is becoming increasingly difficult due to a workforce crisis. The UK currently has a 30% shortfall in radiologists, projected to worsen to 40% by 2028.[2] Reporting backlogs and reliance on outsourcing are escalating, with an estimated 330,000 X-rays in the UK waiting over 30 days for a report.[2, 3]

Autonomous artificial intelligence (AI) reporting of "normal" CXRs, where a device classifies and reports a study as normal without human input, has emerged as a potential solution to reduce pressure on radiology departments.[4] Figure 1 shows the current workflow in comparison to the proposed autonomous AI workflow for CXRs. Several studies show AI may accurately and rapidly distinguish "normal" from "abnormal" CXRs.[5] Autonomous AI could help to alleviate CXR reporting delays and allow radiologists to focus on complex cases, ensuring a quicker turnaround for both patients with normal findings and for those with pathology requiring urgent attention.[4]

**Clinical Performance and Generalisability**

Successful implementation of AI devices requires the ability to perform consistently and equitably across different populations and healthcare systems. The quality and diversity of training and validation datasets are crucial for AI performance. AI devices that are poorly generalisable risk entrenching health inequalities by systematically underperforming in underrepresented groups.[6] Devices must be robustly and prospectively evaluated across all intended populations and healthcare settings to assess real-world accuracy, workflow impact, and technical performance. It is also important to note that errors exist in human reporting. The RCR's CXR reporting standards outline competency expectations[7], and we propose that any autonomous AI tool should demonstrate non-inferiority to such standards.

In clinical settings, diagnostic tools often face a trade-off between sensitivity and specificity. Sensitivity should be prioritised in autonomous deployment to minimise false negatives. Studies report sensitivities above 99% when thresholds are optimised for autonomous reporting, though this has resulted in lower specificities of (28–67%).[6, 8] This low specificity increases false positives, potentially biasing human interpretation and leading to unnecessary investigations. These implications must be considered in pathway design.

**Definition of "Normal"**

The concept of a "normal" CXR lacks standardisation. Interpretations may vary depending on clinical setting and intended use; for instance, should "normal" encompass age-related changes, benign variants, and expected post-surgical findings, or should CXRs be completely "unremarkable". There is no clear consensus on this, and different studies have taken contrasting approaches.[4, 6] Variation in this definition will have a significant impact on the volume reduction possible through autonomous reporting.

Therefore, for autonomous systems to function safely, a clear and consistent definition is essential. This definition should be aligned between developers, clinicians, and regulators. Manufacturers must also disclose the criteria used to define "normal" during model development and validation, to ensure correct interpretation of AI results, meaningful comparison across studies, in addition to legal and regulatory clarity.[4]

**Regulatory Landscape**

Prior to deployment, AI devices must be approved by regulatory authorities who assess their safety, effectiveness, and post-market surveillance strategies. In the United Kingdom, the Medicines and Healthcare Regulatory Agency (MHRA), is responsible for designating approved (UKCA mark) or notified (CE mark) bodies who will assess these tools and award the relevant regulatory approvals. This is done in line with the UK Medical Device Regulation (UK MDR)[9].

Notably, Oxipit was awarded the first CE mark IIb for autonomous reporting of CXR with 'no abnormality', which sets a precent for future regulatory approvals and highlights the need for clear pathways to support safe deployment of autonomous systems in diagnostic workforce.

**Legal and Legislative Considerations**

In the UK, the Ionising Radiation (Medical Exposure) Regulations (IR(ME)R) govern the use of ionising radiation in medical imaging, ensuring that every imaging study has an identified referrer, practitioner, and operator. Currently, a human operator must act as the final signatory for reporting, meaning that implementation of autonomous reporting is currently legally prohibited in the UK[10]. In addition, IR(ME)R defines "reporting" as a recorded clinical evaluation, which includes the outcome and implications; this is dependent on the clinical information and request which AI generated reports may not take into account[10], for instance a "normal" CXR in a lifelong smoker presenting with cough and weight loss may still require follow up CT. We propose that stakeholders should collaborate to understand what safety thresholds and reporting considerations must be met for an amendment of IR(ME)R and to allow the human to be taken out from the system.

Accountability for errors made by AI is another important challenge. Currently, there is no established framework for the accountability of errors made by AI; instructions for use of existing commercial AI devices for radiology typically state that the healthcare professional using the device is responsible for all resulting decisions, but it is difficult to see this apply for autonomous AI. Some have argued that it should be the responsibility of the AI manufacturer, whilst others argue that it would be the hospital trust who has deployed the AI algorithm. Formulation of clear governance procedures and agreement of the legal accountability for errors encountered is paramount. Future case law will undoubtedly play an important role in clarifying liability and establishing precedent for future disputes.

In addition to IR(ME)R, compliance with the UK General Data Protection Regulation presents another challenge. Article 22 grants individuals the right to opt out of decisions made solely through automated processing if those decisions have a legal or similarly significant effect on them, such as a clinical diagnosis[11]. As such, provisions would need to be incorporated into clinical workflows to allow patients to opt out of autonomous reporting, which introduces additional operational complexity and must be addressed in the design of autonomous workflows.

**Ongoing Monitoring and Post-Market Surveillance**

AI models require continuous monitoring to ensure ongoing safety and effectiveness. AI model performance can degrade ("drift") over time due to changes in factors such as patient population demographics, image acquisition hardware, diagnostic criteria and reporting standards[12]. Continuous monitoring of performance in different patient subgroups and clinical environments is important for early recognition and correction of model drift. Updates should be based on data collected from real-world monitoring, to ensure that the algorithm remains relevant and accurate.

Autonomous tools present unique challenges in post-market surveillance. When radiologists are no longer involved in the reporting process, traditional quality assurance methods, such as discrepancy audits, are no longer available. Therefore, mechanisms for monitoring performance and identifying critical incidences must be planned. Reliance on reactive processes such as adverse event reporting is likely to be insufficient given the typically low rate of reporting for medical devices. There should be protocols in place for adverse event monitoring at both a local and national level, to enable identification of any patterns in diagnostic errors made by the AI tool, in addition to any significant patient safety concerns. Regulatory agencies, such as the MHRA, should determine appropriate standards for post market surveillance of autonomous tools on approval to ensure this is upheld by AI manufacturers[9].

**Impact on the Radiology Workforce**

While the removal of "normal" CXRs from radiologist worklists may reduce volume, it could inadvertently increase the complexity of remaining cases. Radiologists may be left with a disproportionate number of complex or subtle findings, increasing diagnostic load, fatigue, and risk of error. There are also concerns around diagnostic calibration. Regular exposure to normal cases helps radiologists maintain confidence in calling a study "normal". If radiologists are only exposed to abnormal cases, this calibration may shift over time, potentially impacting diagnostic thresholds. Additionally, radiology trainees depend on a broad case mix to develop diagnostic proficiency. A reduction in normal case exposure could alter learning opportunities and progression. Strategies such as more frequent breaks or maintaining a mixed case list may mitigate these risks.

**Patient Perspective**

Public trust in autonomous reporting is crucial for adoption. In a recent RCR survey, 80% of respondents supported the use of AI in radiology, yet only 5% endorsed autonomous AI [13]. Ensuring transparency surrounding safeguards in place to prevent AI errors, ongoing monitoring processes, in addition to data protection regulations and privacy, may help to ameliorate patient concerns. Consent processes should be integrated into standard workflows and should explain the use of AI in accessible language. Patients must understand who is responsible for their diagnosis and what options are available in the event of a concern.

**Multi-Professional Stakeholder**

In order to effectively implement autonomous reporting in the UK, we propose that a multi-stakeholder group should be established, involving the Royal College of Radiologists, College of Radiographers, industry stakeholders, regulators, patients, and clinicians across specialties who rely on imaging to guide care. As autonomous reporting removes the safety net of a human report, it is vital to assess acceptability across all professional groups affected and ensure guidance reflects the realities of multidisciplinary clinical practice.

**Conclusion**

Autonomous reporting of normal CXRs has the potential to revolutionise healthcare by improving radiology efficiency in an era of workforce shortages. Figure 2 illustrates the key strengths, weaknesses, opportunities, and threats associated with this approach. Therefore, for autonomous AI to become a reality, significant steps must be taken to validate AI performance, meet regulatory requirements, clarify legal frameworks, and maintain ongoing oversight. Performance benchmarks and post-market surveillance strategies must be rigorous to ensure the highest level of patient care, and addressing the concerns of radiologists and patients will be vital for successful and responsible implementation.

**Figure Captions:**

**Figure 1 – Reporting workflow for chest X-ray reporting at present, in comparison to a proposed reporting workflow with autonomous artificial intelligence deployed**

**Figure 2 – Strengths, weaknesses, opportunities and threats (SWOT) analysis for autonomous reporting**

Figure 1

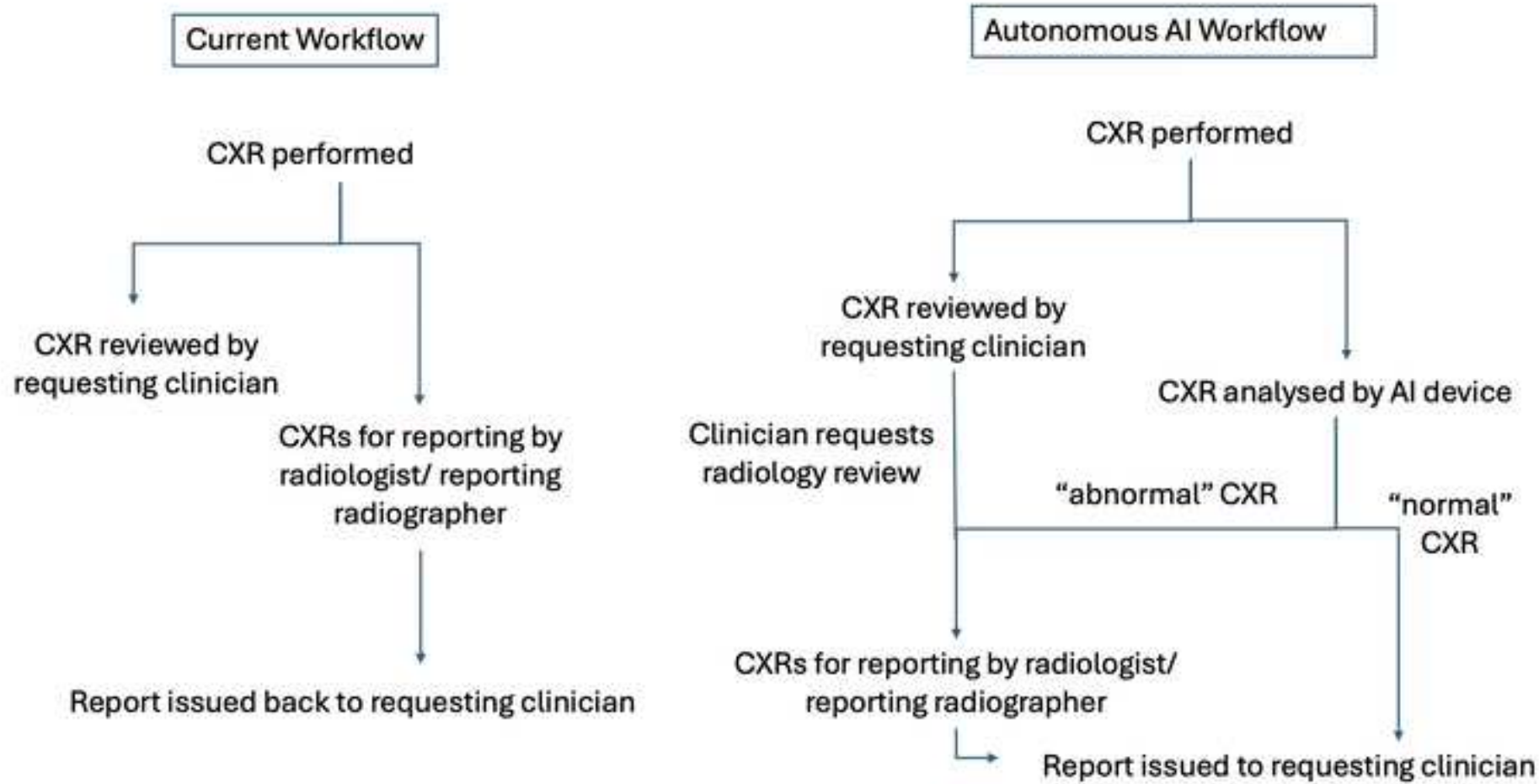

**Strengths**
- Reduces radiologist workload and NHS backlogs
- Faster reporting for patients with normal findings
- High sensitivity (>99%) in optimised models, meaning few false negatives
- Allows radiologists to focus on complex/pathological cases
- Addresses NHS radiologist shortfall (30-40%)
- Proven feasibility (CE mark IIb for Oxipit)

**Weaknesses**
- No standardised definition of a "normal" CXR
- Current UK law (IR(ME)R) prohibits full autonomous
- Low specificity may increase false positives
- No clear legal framework for accountability for errors
- Potential loss of diagnostic calibration and training exposure
- No defined post-market surveillance methodology
- GDPR opt-out rights introduce workflow complexity

**Opportunities**
- Position UK as leader in safe, regulated AI use
- Reduce dependence on outsourcing and locums, leading to a potential cost-saving
- Improve NHS digital infrastructure and workflows
- Enhance patient experience via a quicker turnaround

**Threats**
- Public mistrust or low confidence in autonomous AI
- Risk of bias or underperformance in certain populations
- Risk of AI performance "drift" without monitoring
- Legal ambiguity over who is liable for AI driven errors
- Resistance from clinicians fearing replacement of de-skilling